\documentclass[sigconf]{acmart}

\setcopyright{none} 

\usepackage{multirow}
\usepackage{marvosym}
\usepackage{tabu}
\usepackage{balance}
\usepackage{amssymb}
\usepackage{amscd}
\usepackage{amsmath}
\usepackage{xcolor}
\usepackage{listings}
\usepackage{soul}
\usepackage{xspace}

\usepackage{textcomp,booktabs}
\usepackage{colortbl}
\definecolor{mygray}{gray}{.9}

\usepackage{multirow}
\usepackage{graphicx}
\usepackage{subfigure}
\usepackage{hyperref}

\usepackage[linesnumbered,ruled]{algorithm2e}

\SetCommentSty{mycommfont}

\usepackage{tcolorbox}
\tcbuselibrary{breakable, skins}

\tcbset{%
	label begin/.style={label={#1}},
	label end/.style={after upper=\label{#1}}
}

\newtcolorbox[%
auto counter]{mybox}[2][]{%
	enhanced jigsaw,
	breakable,
	#1}

\newcommand{\ignore}[1]{}
\newcommand{\revised}[1]{}

\begin{document}
	\title{Understanding Android Obfuscation Techniques: \\ A Large-Scale Investigation in the Wild}

\author{Shuaike Dong$^{1}$, Menghao Li$^{2}$, Wenrui Diao$^{3}$, Xiangyu Liu$^{4}$, Jian Liu$^{2}$, Zhou Li$^{5}$, Fenghao Xu$^{1}$, \newline Kai Chen$^{2}$, XiaoFeng Wang$^{6}$, and Kehuan Zhang$^{1}$}
\affiliation{%
	\institution{$^{1}$The Chinese University of Hong Kong, Email: \{ds016, xf016, khzhang\}@ie.cuhk.edu.hk}
}
\affiliation{%
	\institution{$^{2}$Institute of Information Engineering, Chinese Academy of Sciences, Email: \{limenghao, liujian6, chenkai\}@iie.ac.cn}
}
\affiliation{%
	\institution{$^{3}$Jinan University, Email: diaowenrui@link.cuhk.edu.hk}
}
\affiliation{%
	\institution{$^{4}$Alibaba Inc., Email: eason.lxy@alibaba-inc.com}
}
\affiliation{%
	\institution{$^{5}$ACM Member, Email: lzcarl@gmail.com}
}
\affiliation{%
	\institution{$^{6}$Indiana University Bloomington, Email: xw7@indiana.edu}
}

\begin{abstract}
	Program code is a precious asset to its owner. Due to the easy-to-reverse nature of Java, code protection for Android apps is of particular importance. To this end, code obfuscation is widely utilized by both legitimate app developers and malware authors, which complicates the representation of source code or machine code in order to hinder the manual investigation and code analysis. Despite many previous studies focusing on the obfuscation techniques, however, our knowledge on how obfuscation is applied by real-world developers is still limited.
	
	In this paper, we seek to better understand Android obfuscation and depict a holistic view of the usage of obfuscation through a large-scale investigation in the wild. In particular, we focus on four popular obfuscation approaches: identifier renaming, string encryption, Java reflection, and packing. To obtain the meaningful statistical results, we designed efficient and lightweight detection models for each obfuscation technique and applied them to our massive APK datasets (collected from Google Play, multiple third-party markets, and malware databases). We have learned several interesting facts from the result. For example, malware authors use string encryption more frequently, and more apps on third-party markets than Google Play are packed. We are also interested in the explanation of each finding. Therefore we carry out in-depth code analysis on some Android apps after sampling. We believe our study will help developers select the most suitable obfuscation approach, and in the meantime help researchers improve code analysis systems in the right direction.
	
\end{abstract}
	

	

\maketitle

\section{Introduction}
\label{sec:intro}

Code is a very important intellectual property to its developers, no matter if they work as individuals or for a large corporation. To protect this property, \textit{obfuscation} is frequently used by developers, which is also considered as a double-edged sword by the security community. To a legitimate software company, obfuscation keeps its competitors away from copying the code and quickly building their own products in an unfair way. To a malware author, obfuscation raises the bar for automated code analysis and manual investigation, two approaches adopted by nearly every security company. For a mobile app, especially the one targeting Android platform, obfuscation is particularly useful, given that the task of disassembling or decompiling Android app is substantially easier than doing so for other sorts of binary code, like X86 executables. 

Android obfuscation arguably is pervasive. On the one hand, there are already more than 2.8 million apps available for downloading just in one app market, Google Play, up to March 2017~\cite{url_googleplay}. On the other hand, many off-the-shelf obfuscators are developed, and some authors claim their tools are used by more than 300,000 apps ~\cite{url_360jiagu}. Consequently, the issues around app obfuscation attract many researchers. So far, most of the studies focus on the topics like what obfuscation techniques can be used~\cite{apvrille2014obfuscation}, how they can be improved~\cite{DBLP:conf/dasc/ShuLZG14}, how well they can be handled by state-of-art code analysis tools~\cite{DBLP:conf/ccs/RastogiCJ13}, and how to deobfuscate the code automatically~\cite{DBLP:conf/ccs/BichselRTV16}. While these studies provide solid ground for understanding the obfuscation ~\textit{techniques} and its ~\textit{implications}, there is a still an unfilled gap in this domain: how obfuscation is \textit{actually used} by the vast amount of developers?

We believe this topic needs to be studied, and the answer could enlighten new research opportunities. To name a few, for developers, learning which obfuscation techniques should be used is quite important. Not all obfuscation techniques are equally effective, and using some might even bring the incompatibility issue. Plenty of code analysis approaches were proposed, but their effects are usually hampered by obfuscation and the impact greatly differs based on the specific obfuscation technique in use, e.g., identifier renaming is much less of an issue comparing to string encryption. Knowing the distribution of obfuscation techniques can better assist the design of code analysis tools and prioritize the challenges need to be tackled. All roads paving to the correct conclusions call for measurement on real-world apps, and only the result coming from a comprehensive study covering a diverse portfolio of apps (published in different markets, in different countries, from both malware authors and legitimate companies) is meaningful.

\vspace{2pt} \noindent
\textbf{Our Work}. As the first step, in this paper, we systematically study the obfuscation techniques used in Android apps and carry out a large-scale investigation for apps in the wild. We focus on four most popular Android obfuscation techniques (identifier renaming, string encryption, Java reflection, and packing) and measure the base and popular implementation of each technique. To notice, the existing tools, like deobfuscators, cannot solve our problem here, since they either work well against a specific technique or a specific off-the-shelf obfuscator (e.g., ProGuard). As such, they cannot be used to provide a holistic view. Our key insight to this end is that instead of mapping the obfuscated code to its original version, a challenge not yet fully addressed, we only need to \textit{cluster} them based on their code patterns or statistical features. Therefore, we built a set of lightweight detectors for all studied techniques, based on machine learning and signature matching. Our tools are quite effective and efficient, suggested by the validation result on ground-truth datasets. We then applied them on a real-world APK dataset with 114,560 apps coming from three different sources, including Google Play set, third-party markets set, and malware set, for the large-scale study.

\vspace{2pt} \noindent
\textbf{Discoveries}. Our study reveals several interesting facts, with some confirming people's intuition but some contradicting to common beliefs: for example, as an obfuscation approach, identifier renaming is more widely-used in third-party apps than in malware. Also, though basic obfuscation is prevalently applied in benign apps, the utilization rate of other advanced obfuscation techniques is much lower than that of malware. We believe these insights coming from ``\textit{big code}'' are valuable in guiding developers and researchers in building, counteracting or using obfuscation techniques. 

\vspace{2pt} \noindent 
\textbf{Contributions.} We summarize this paper's contributions as below:

\begin{itemize}
	\item \textbf{Systematic Study}. We systematically study the current mainstream Android obfuscation techniques used by app developers.
	
	\item \textbf{New Techniques}. We propose several techniques for detecting different obfuscation techniques accurately, such as n-gram -based renaming detection model and backward slicing-based reflection detection algorithm.  
	
	\item \textbf{Large-scale Evaluation}. We carried out large-scale experiments and applied our detection techniques on over 100K APK files collected from three different sources. We listed our findings and provided explanations based on in-depth analysis of obfuscated code.
\end{itemize}

\vspace{2pt} \noindent 
\textbf{Roadmap}. The rest of this paper is organized as follows: We systematically summarize popular Android obfuscation techniques in Section~\ref{sec:background}. Section~\ref{sec:system} overviews the high-level architecture of our detection framework. The detailed detection strategies and statistical results on large-scale datasets are provided in Section~\ref{sec:findings}. Also, we discuss some limitations and future plans in Section~\ref{sec:discussion}. Section~\ref{sec:related} reviews the previous research on Android obfuscation, and Section~\ref{sec:conclusion} concludes this paper.

\section{Background}
\label{sec:background}

In this section, we briefly introduce the structure of APK file and overview some common Android obfuscation techniques.

\subsection{APK File Structure}
An APK (Android application package) file is a zip compressed file containing all the content of an Android app, in general, including four directories (\texttt{res}, \texttt{assets}, \texttt{lib}, and \texttt{META-INF}) and three files (\texttt{AndroidManifest.xml}, \texttt{classes.dex}, and \texttt{resources.arsc}). The purposes of these directories and files are listed as below.

\begin{description}
	\item[\texttt{res}] This directory stores Android resource files which will be mapped into the \texttt{.R} file in Android and allocated the corresponding ID.
	
	\item[\texttt{assets}] This directory is similar to the \texttt{res} directory and used to store static files in the APK. However, unlike \texttt{res} directory, developers can create subdirectories in any depth with the arbitrary file structure.
	
	\item[\texttt{lib}] The code compiled for specific platforms (usually library files, like \texttt{.so}) are stored in this directory. Subdirectories can be created according to the type of processors, like armeabi, armeabi-v7a, x86, x86\_64, mips.
	
	\item[\texttt{META-INF}] This directory is responsible for saving the signature information of a specific app, which is used to validate the integrity of an APK file.
	
	\item[\texttt{AndroidManifest.xml}] This XML file is the configuration of an APK, declaring its basic information, like name, version, required permissions and components. Each APK has an AndroidManifest file, and the only one.
	
	\item[\texttt{classes.dex}] The dex file contains all the information of the classes in an app. The data is organized in a way the Dalvik virtual machine can understand and execute.
	
	\item[\texttt{resources.arsc}] This file is used to record the relationship between the resource files and related resource ID and can be leveraged to locate specific resources.
\end{description}

\subsection{Android Obfuscation Characterization}
In general, obfuscation attempts to garble a program and makes the source or machine code more difficult for humans to understand. Programmers can deliberately obfuscate code to conceal its purpose or logic, in order to prevent tampering, deter reverse engineering, or behave as a puzzle for someone reading the code. Specifically, there are several common obfuscation techniques used by Android apps, including identifier renaming, string encryption, excessive overloading, reflection, and so forth.

\vspace{2pt} \noindent 
\textbf{Identifier Renaming.} In software development, for good readability, code identifiers' names are usually meaningful, though developers may follow different naming rules (like CamelCase, Hungarian Notation). However, these meaningful names also accommodate reverse-engineers to understand the code logic and locate the target functions rapidly. Therefore, to reduce the potential information leakage, the identifier's name could be replaced by a meaningless string. The following code snippet gives an example, in which all identifiers in class \texttt{Account} are renamed.

{\small
	\begin{lstlisting}[
	language={Java},
	label={lst:xxx},
	keywordstyle=\color{blue!70},
    numbers=left,                    
	numbersep=5pt, 
	xleftmargin=8pt,
	xrightmargin=5pt,
	numberstyle=\scriptsize\color{gray},
	breaklines=true,   
	frame=single,
	%frame=shadowbox,
	basicstyle=\ttfamily,
	commentstyle=\color{blue} \textit,
	stringstyle=\itfamily,
	showstringspaces=false]
public class a{
    private Integer a;
    private Float = b;
    public void a(Integer a, Float b){
    	this.a = a + Integer.valueOf(b)
    }
}
	\end{lstlisting} 
}

\vspace{2pt} \noindent 
\textbf{String Encryption.} Strings are very common-used data structures in software development. In an obfuscated app, strings could be encrypted to prevent information leakage. Based on cryptographic functions, the original plaintexts are replaced by random strings and restore at runtime. As a result, string encryption could effectively hinder \emph{hard-coded} static scanning. The following code block shows an example.

{\small
	\begin{lstlisting}[
	language={Java},
    label={lst:xxx},
	keywordstyle=\color{blue!70},
	numbers=left,                    
	numbersep=5pt, 
	xleftmargin=8pt,
	xrightmargin=5pt,
    numberstyle=\scriptsize\color{gray},
	breaklines=true,
	frame=single,
	basicstyle=\ttfamily, 
	commentstyle=\color{blue} \textit,
	stringstyle=\ttfamily, 
	showstringspaces=false,
	mathescape=true]
String option = "@^@#\x `1 m*7 %**9_!v";
this.execute($\textbf{decrypt}$(option));
	\end{lstlisting} 
}




\vspace{2pt} \noindent 
\textbf{Java Reflection.} Reflection is an advanced feature of  Java~\cite{url_reflection}, which provides developers with a flexible approach to interact with the program, e.g., creating new object instances and invoking methods dynamically. One common legitimate usage is to invoke nonpublic APIs in the SDK (with the annotation \texttt{@hide}). The following code block gives an example of reflection that invokes a hidden API \texttt{batteryinfo}.

{\small
	\begin{lstlisting}[
	language={Java},
    label={lst:reflection},
	keywordstyle=\color{blue!70},
	numbers=left,                    
	numbersep=5pt, 
	xleftmargin=8pt,
	xrightmargin=5pt,
	numberstyle=\scriptsize\color{gray},
	breaklines=true,
	frame=single,
	%frame=shadowbox,
	basicstyle=\ttfamily,
	commentstyle=\color{blue} \textit,
	stringstyle=\ttfamily,
	showstringspaces=false]
Object object = new Object();
Method getService = Class.forName("android.os.ServiceManager").getMethod("getService", String.class);
Object obj = getService.invoke(object, new Object[]{new String("batteryinfo")});
	\end{lstlisting} 
}

As an obfuscation technique, reflection is a good choice of hiding program behaviors because it can transfer the control to a certain function implicitly, which can not be well handled by state-of-the-art static analysis tools. Therefore, malware developers usually heavily employ reflection to hide malicious actions.

\vspace{2pt} \noindent 
\textbf{Packing.} Packing is a widely-used code protection technique. The packed APK file is composed of an encrypted origin APK and a wrapper APK. When the user launches the APK, the wrapper will run first, decrypt the original APK and load it into the memory, and then the execution will be handed to the decrypted APK. Due to the cryptographic procedure and runtime release, it becomes hard to get the original code through static analysis. We regard packing as an obfuscation skill in a broad sense because its goal is to hinder the reverse-engineering as well.


\section{System Design}
\label{sec:system}

Our target is to systematically study the Android obfuscation techniques and carry out a large-scale investigation. As the first step, we design an efficient Android code analysis framework to identify the obfuscation techniques used by developers. Here we overview the high-level design of this framework and introduce the datasets prepared for the subsequent large-scale investigations.

\subsection{System Overview}

\begin{figure*}[t]
	\centering
	\includegraphics[width=1\textwidth]{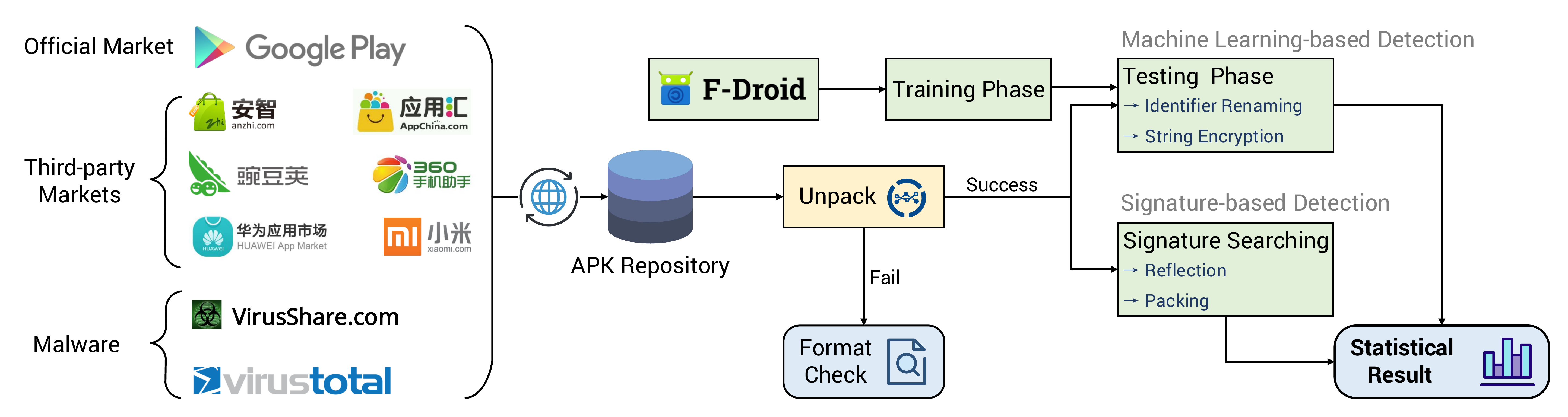}           
	\caption{Android App Obfuscation Detection Framework}
	\label{fig:framework}
\end{figure*}

To detect the usage of obfuscation techniques, we propose an architecture to analyze APK files automatically, as illustrated in Figure~\ref{fig:framework}. After the APK files collected from several channels (details are provided in Section~\ref{subsec:dataset}) are stored in our server, this detection framework will try to unpack them for the primary testing. Some damaged APK files failing to pass this step will be discarded. Then this framework applies four targeted detection methods to identify obfuscated Smali code blocks. These detection methods could be classified into two categories: signature-based and machine learning-based. For the obfuscation techniques with specific features, we search the corresponding signatures in Smali code to determine the existence. For example, the reflective calls which implicitly invoke another function can be located by searching the sequence pattern [\texttt{Class.forName()}$\rightarrow$\texttt{getMethod()}$\rightarrow$\texttt{invoke()}]. However, it is difficult to extract fixed features for some techniques (e.g., encrypted strings), so we utilize machine learning algorithms to classify automatically. The training set comes from F-Droid~\cite{url_fdroid}, an open source Android app repository.

\subsection{APK Dataset}
\label{subsec:dataset}
We are interested in the obfuscation usage status of apps in different types, so three representative APK datasets were used in our experiment: Google Play set (26,614 samples), third-party market set (65,666 samples), and malware set (22,280 samples). These samples were collected during 2016 and 2017. In total, our experiment dataset contains 114,560 sample with the size of around 1.521TB. More details are given in Table~\ref{tab:datset}.

As the official app store for Android, Google Play is the main Android app distribution channel. Thus, its sample set could reflect the deployment status of obfuscation used by mainstream developers. Also, due to the policy restriction, in some countries (such as China), Google Play is not available, and users have to install apps from third-party markets. Therefore, in the second dataset, we select six popular app markets from China (say Anzhi~\cite{url_anzhi}, Xiaomi~\cite{url_xiaomi}, Wandoujia~\cite{url_wandoujia}, 360~\cite{url_360}, Huawei~\cite{url_huawei}, and AppChina~\cite{url_appchina}) and developed the corresponding crawlers to collect their apps. Note that the replicated samples from different markets have been excluded. Lastly, except for legitimate app samples, we are also curious about whether malware authors heavily use obfuscation skills to hide their malicious intentions. So, the last dataset contains the malware samples coming from VirusShare~\cite{url_virusshare} and VirusTotal~\cite{DBLP:conf/bigdataconf/HuangZZZZ0CZ16, url_virustotal}.

\begin{table}[t]
	\centering
	\caption{APK Dataset for Investigation}
	\label{tab:datset}
	
	\begin{tabu} to 0.96\columnwidth{|X[2.5,c]|X[2.5,c]|X[2.5,c]|}
		\hline
		\textbf{Type} & \textbf{Source} & \textbf{Number} \\
		\hline
		\hline
		\textbf{Official Market} & Google Play & 26,614 \\
		\hline
		\multirow{6}{*}{\textbf{3rd-party Market}} &
		Wandoujia & 8,979 \\
		\cline{2-3}
		& 360 & 18,724 \\
		\cline{2-3}
		& Huawei & 22,048 \\
		\cline{2-3}
		& Anzhi & 7,121 \\
		\cline{2-3}
		& Xiaomi & 4,649 \\
		\cline{2-3}
		& AppChina & 4,145\\
		\cline{1-3}
		\multirow{2}{*}{\textbf{Malware}} &
		VirusShare & 19,004 \\
		\cline{2-3}
		& VirusTotal & 3,267 \\
		\cline{1-3}
		
	\end{tabu}    
\end{table}

\section{Obfuscation Detections and Large-Scale Investigation}
\label{sec:findings}

In this section, we introduce the detection approaches for each obfuscation technique and summarize our findings based on large-scale experiments. 

\subsection{Identifier Renaming}
\label{subsec:identifier_renaming}

Generally, in the software development, the names of identifiers (variable names, function names, and so forth) are usually meaningful, which could provide good code readability and maintainability. However, such \emph{clear} names may leak much information due to the easy-to-reverse feature of Java. As a solution, identifier renaming is proposed and widely used in practice.

The renaming operation can be appended at different stages of APK file packaging. For example, ProGuard~\cite{url_proguard} and Allatori~\cite{url_allatori} work at the source-code level, mapping the original names to mangled ones based on the user's configuration. The other obfuscators, like DashO~\cite{url_dasho}, DexProtector\cite{url_dexprotector}, and Shield4J~\cite{url_dshield4J}, can work directly on APK files, modifying \texttt{.class} and \texttt{.dex} files.

Given an identifier, we can easily tell whether some obfuscator has renamed it based on the information it contains. In other words, if an identifier name is obscure and meaningless, it can be regarded as obfuscated because it tries to hide the actual purpose and intention. A typical renaming operation is changing the original name to a single character (like "\textsf{a}", "\textsf{b}") or some kind of puzzling string (like "\textsf{IlllIlII}", "\textsf{oO00O0oo}")~\cite{apvrille2014obfuscation}. However, the manual check is obviously not qualified for our large-scale scanning goal. Moreover, we focus on the whole APK contents rather than a single identifier. Therefore, we need to design a robust and systematic detection method for identifier renaming.

Beyond that, as a special case of identifier renaming, the \emph{excessive overloading} technique utilizes the overloading feature of Java and could map irrelevant identifier names to the same one, making the code more confusing to analysts~\cite{DBLP:journals/compsec/BalachandranSTT16}. For example, in the sample \texttt{idfhn}\footnote{MD5: \texttt{7d9eb791c09b9998336ef00bf6d43387}}, more than 46 functions are named as \texttt{idfhn} (the same as the package name). Though the compiler could distinguish these variables with the same name, security analysts have to face more troubles. In our research, we also paid attention to the application of overloading feature and its impact on code analysis.



\vspace{2pt} \noindent
\textbf{Identifier Renaming Detection.} To the above challenges and targets, we combine the computational linguistics and machine learning techniques for accurate renaming detection. The high-level idea is based on the probabilistic language model. The insight is that identifier renaming will lead to the abnormal distribution of characters and character combinations, which can be used to distinguish from normal ones (non-obfuscated). Here we give our three-step approach:
 
	
	
	
	

\begin{enumerate}
    \item \textit{Data Pre-processing.} As the most frequently used three identifiers, the names of all classes, methods, and fields of the target APK sample are extracted as the training candidates. Note that, software developers often introduce third-party libraries into their apps instead of redevelopment. However, those third-party libraries may also contain obfuscated code, which can not reflect the protection deployed by developers proactively.\ignore{which may affect our obfuscation utilization detection.} Therefore, we have pre-removed over 12,000 common third-party libraries using the approach of Li et al.~\cite{lilibd}.

    \item \textit{Feature Generation.} The amount of identifiers varies among different apps. To build a uniform expression, we apply the $n$-gram algorithm~\cite{url_n-gram} to generate a fixed-length feature vector for each app. An $n$-gram is a contiguous sequence of $n$ items from a given sequence of text or speech. In our implementation, we apply $3$-gram\footnote{For example, if there is a string "\textsf{abcdefgh}", all of the 3-gram sequences it contains are \{\textsf{abc}, \textsf{bcd}, \textsf{cde}, \textsf{def}, \textsf{efg}, \textsf{fgh}\}.} to traverse each name string in extracted raw name set to form a fixed-length\footnote{The length is restricted by the legal characters sets used for contracting a name in Java: ["\textsf{a-z}", "\textsf{ A-Z}", "\textsf{0-9}", "\textsf{\_}", "\textsf{\$}", "\textsf{$\backslash$}"].} feature vector. The feature vector records the frequency of each three continuous characters and will be normalized. 
    
    \item \textit{Classification.} The training set is based on an open-source Android app repository -- F-Droid~\cite{url_fdroid}. We apply different obfuscators on these Android source code to generate obfuscated apps as the ground truth. Lastly, we choose Support Vector Machine (SVM) as the classification algorithm. 
\end{enumerate}

\vspace{2pt} \noindent
\textbf{Experiment Settings.} We implemented a prototype of our detection model based on Androguard~\cite{url_androguard} with more than 1,500 Python lines of code. For training, we downloaded 3,147 apps and their corresponding source code from F-Droid. Two obfuscators, ProGuard and DashO, were used to generate the obfuscated samples because they have different renaming policies. Note that, due to the diversity of apps' project configurations, not all of them can be processed by both ProGuard (2,107 successful samples) and DashO (654 successful samples). Among them, we randomly chose 500 original apps and 500 obfuscated apps (250 for ProGuard and 250 for DashO) as the training set. 

\ignore{To validate the effectiveness of our renaming detection model, we constructed a small-scale testing dataset (250 original apps and 250 obfuscated apps), also based on F-Droid. The detection result is quite satisfactory, say 95.4\% precision rate with false positive (FP) 5.8\% and false negative (FN) 0.6\%.}

We then conducted three steps to validate the effectiveness of our renaming detection model. First, we randomly selected 1,000 original apps and did manual check to make sure that they were non-obfuscated. Our classifier completely correctly labeled these apps as "original", which means the false positive rate is 0\%. We then tested our model on 1,000 obfuscated apps(500 obfuscated by Proguard and 500 by DashO) and our model mis-classified 6 samples(5 from Proguard, 1 from DashO), reaching a 0.6\% false negative rate in total. Due to identifier renaming will lead to an abnormal distribution of character combinations, we consider our model can be generalized to other obfuscators even if they have different implementing policies. To verify this, we conducted a third experiment. We collected another testing set consisting of 200 samples obfuscated by another obfuscator Allatori. The completely successful classification results showed our model's good attribute of generalization. 

\vspace{2pt} \noindent
\textbf{Large-scale Investigation and Findings.} The purpose of our study is to plot the current usage status of Android obfuscation in the wild. Therefore, we carried out a large-scale detection on the three typical datasets (Google Play, third-party markets, and malware) mentioned in Section~\ref{subsec:dataset}. The obfuscation detection result by dataset is given in Figure~\ref{fig:renaming_percentage}. According to such statistics, we have two immediate findings:

\begin{figure}[t]
	\centering
	\includegraphics[width=0.95\columnwidth]{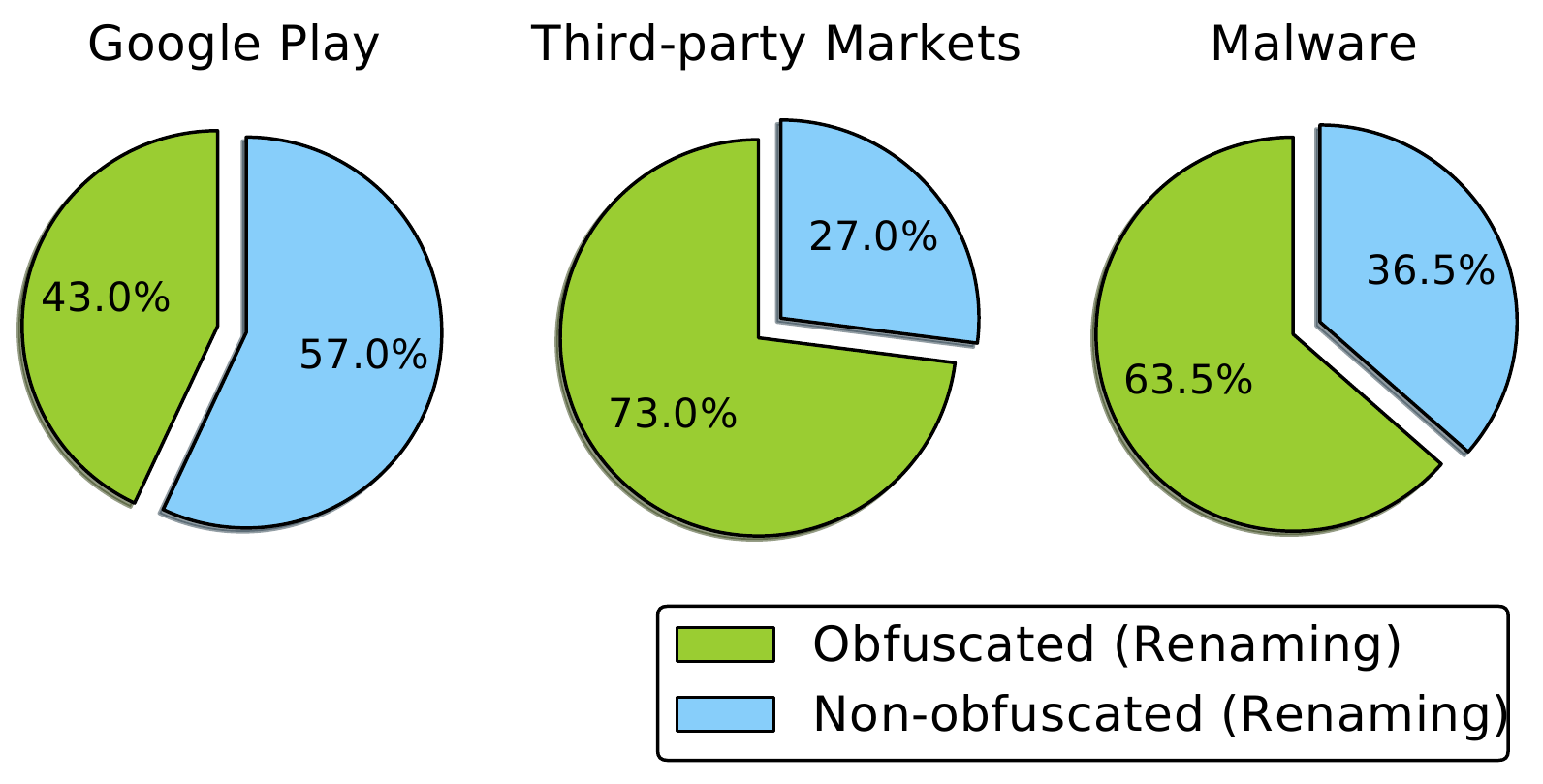}           
	\caption{Ratio of Identifier Renaming in Three Datasets}
	\label{fig:renaming_percentage}
\end{figure}

\begin{mybox}[boxsep=0pt,
	boxrule=1pt,
	left=4pt,
	right=4pt,
	top=4pt,
	bottom=4pt,
	]
	 ~$\Rightarrow$ 1. Compared with the apps on Google Play, the ones from third-party markets apply more renaming operations.\\
     $\Rightarrow$ 2. Over one third of malware don't apply identifier renaming.
\end{mybox}

To the first finding, we ascribe it to the discrepancy between app market environments. The piracy issue in Chinese app markets are quite severe~\cite{url_chinesepiracy}, say nearly 20\% apps are repacked or cloned~\cite{DBLP:conf/icse/ChenLZ14}. Such situation urges developers to put more effort into protecting their apps. On the other hand, Google Play provides more strict and timely supervision, which mitigates the severity of software piracy largely. The better application ecosystem makes many developers believe obfuscation is just an optional protection approach.

To the second finding, the percentage of malware utilizing identifier renaming is only 63.5\%, slightly less than third-party apps, which is opposite our traditional opinion. After manually checking the code of malware without renaming-obfuscation, we conclude that two aspects contribute to such phenomenon.

\begin{itemize}
	\item \emph{Script Kiddies}. Many entry-level malware authors only could develop simple malicious apps and lack the knowledge of how to disguise malicious behaviors through obfuscation. 
	
	\item \emph{False Alarmed "Malware"}. For some apps, their main bodies are benign and non-obfuscated, while the imported third-party libraries contain some kinds of sensitive and suspicious behaviors which are recognized as malicious by some anti-virus software. A common example is the advertising library.
\end{itemize}




\vspace{2pt}
In addition, we explored the difference in renaming implementation between malware and benign apps. The result reflects: 
\begin{mybox}[boxsep=0pt,
	boxrule=1pt,
	left=4pt,
	right=4pt,
	top=4pt,
	bottom=4pt,
	]
	~ $\Rightarrow$ Malware authors prefer to use more complex renaming policies.
\end{mybox}

We find that, in benign apps (the samples on Google Play and third-party markets), most identifier names are mapped to \{\textsf{a}, \textsf{b}, \textsf{aa}, \textsf{ab}, \textsf{aaa}, \textsf{$\dots$}\} and so on, in lexicographic order. In fact, such renaming rules accord with the default configurations of many obfuscators (such as ProGuard). That is to say, app developers do not intend to change the renaming rules to more ingenious ones. However, malware authors usually put more effort into configuring the renaming policies. For example, some malware samples utilize special characters (encoded in Unicode) as obfuscated names (e.g., \`{E}, \^{o}), which seems very odd but still be regarded as legal by Java compilers. Also, some dazzling weird names (like \{\textsf{IlllIlII}, \textsf{oO00O0oo}, \textsf{$\dots$}\}) could be found.

\vspace{2pt}
Based on the result of excessive overloading detection, we find: 
\begin{mybox}[boxsep=0pt,
	boxrule=1pt,
	left=4pt,
	right=4pt,
	top=4pt,
	bottom=4pt,
	]
    ~$\Rightarrow$ 1. The deployment rate of excessive overloading approximates that of identifier renaming. \\
    ~$\Rightarrow$ 2. Malware may use irrelevant names to hide the true intention.
\end{mybox}

Our statistics show that most of the excessive overloading cases appear along with identifier renaming. The reason may derive from that many obfuscators configure the excessive overloading by default. For example, Proguard provides the option "\texttt{-overloadaggressively}" for convenient deployment.

To the second finding, we find there are also some non-name-obfuscated samples applying overloading to confuse analysts. In sample \texttt{tw.org.ncsist.mdm}\footnote{MD5: \texttt{01a93f7e94531e067310c1ee0f083c07}}, the name of overloaded function \texttt{attachBaseContext} (A \texttt{protected} method in class \texttt{android.app. Application}) will mislead security analysts because the logic of this function is implemented for encryption.

\subsection{String Encryption}
The strings in a \texttt{.dex} (Dalvik executable) file may leak a lot of private information about the program. As security protection, those hard-coded texts can be stored in an encrypted form to prevent reverse analysis. In this section, we take a deep insight into the string encryption and focus on two aspects:

\begin{enumerate}
	\item Detect whether an app uses the string encryption.
	\item Analyze the cryptographic functions invoked by apps.
\end{enumerate}

\vspace{2pt} \noindent
\textbf{String Encryption Detection.} Similar to the approach for identifier renaming detection (Section~\ref{subsec:identifier_renaming}), we trained a machine-learning based model to classify encrypted strings and plain-text strings. We reused the $n$-gram algorithm, SVM algorithm, and the open-source apps from F-Droid. Here we only describe the different steps. At first, all strings appeared in an app are extracted. Next, a vector was generated for each app via 3-gram algorithm. Distinct from the setting for identifier renaming detection, there is no restriction on the content of a string. Therefore, we extended the acceptable character set to all ASCII codes.

In the implementation, we reused most code of identifier renaming detection model. Since string encryption is not a common function provided by off-the-shelf obfuscators, we chose DashO and DexProtector to generate the ground truth and finally obtained 737 string-encrypted samples for training. To avoid the overfitting caused by unbalanced data, we randomly selected 500 original apps and 500 string-encrypted apps to train our model. To verify the effectiveness, we randomly selected another 100 original apps and 100 string-encrypted apps for testing. The result shows our model could achieve 98.5\% success rate with FP 1\% and FN 2\%. 

\vspace{2pt} \noindent
\textbf{Cryptographic Function Analysis.} Previous work has proposed various approaches to identify cryptographic functions in a program, like~\cite{DBLP:conf/ccs/CalvetFM12, DBLP:conf/raid/GrobertWH11, DBLP:conf/malware/MatenaarWLG12}. Those methods were specifically designed for the identification of the standard, modern cryptographic algorithms in binary code, like AES, DES, and RC4. The features used by the previous commonly include entropy analysis, searchable constant patterns, excessive use of bitwise arithmetic operations, memory fetch patterns and so on, besides, the dynamic binary instrument is also widely-used by analysts to better locate and identify the cryptographic primitives. However, previous approaches do not fit android platform very well due to three reasons: (1) Smali instructions have different representations from the x86 assembly language, especially for memory access. (2) Java provides the complete implementations of standard cryptographic algorithms through Java Cryptography Extension~\cite{JAE_url}. Therefore, in most cases, developers do not need to implement cryptographic related functions again. (3) Java provides a series of string \& character operations, like \texttt{concat()}, \texttt{substring()}, \texttt{getChars()}, \texttt{strim()} and so on, which can be used to build an encrypted string. 

To better handle the identification in Android apps, we extended the previous approaches with more empirical features, shown as below.

\ignore{
Noted that for string-encrypted apps, strings will be restored to plain forms when used, such as delivered as the parameter of another function. The decoding procedure is commonly completed by decryption functions, leading to its higher frequency of appearing with encrypted strings, which can be used as a statistical feature.
}

\begin{itemize}
\item The ratio of bit and loop operations.
\item The usage of Java Cryptography Extension API invoking.
\item The amount of operations on string \& character variables.
\item The frequency of encrypted strings as function parameters (for decryption function).
\end{itemize}

\vspace{2pt} \noindent
\textbf{Large-scale Investigation and Findings.} We applied our string encryption detection model on the testing datasets. The results are presented in Figure~\ref{fig:encryption_percentage}. The direct findings are that:

\begin{figure}[t]
	\centering
	\includegraphics[width=0.95\columnwidth]{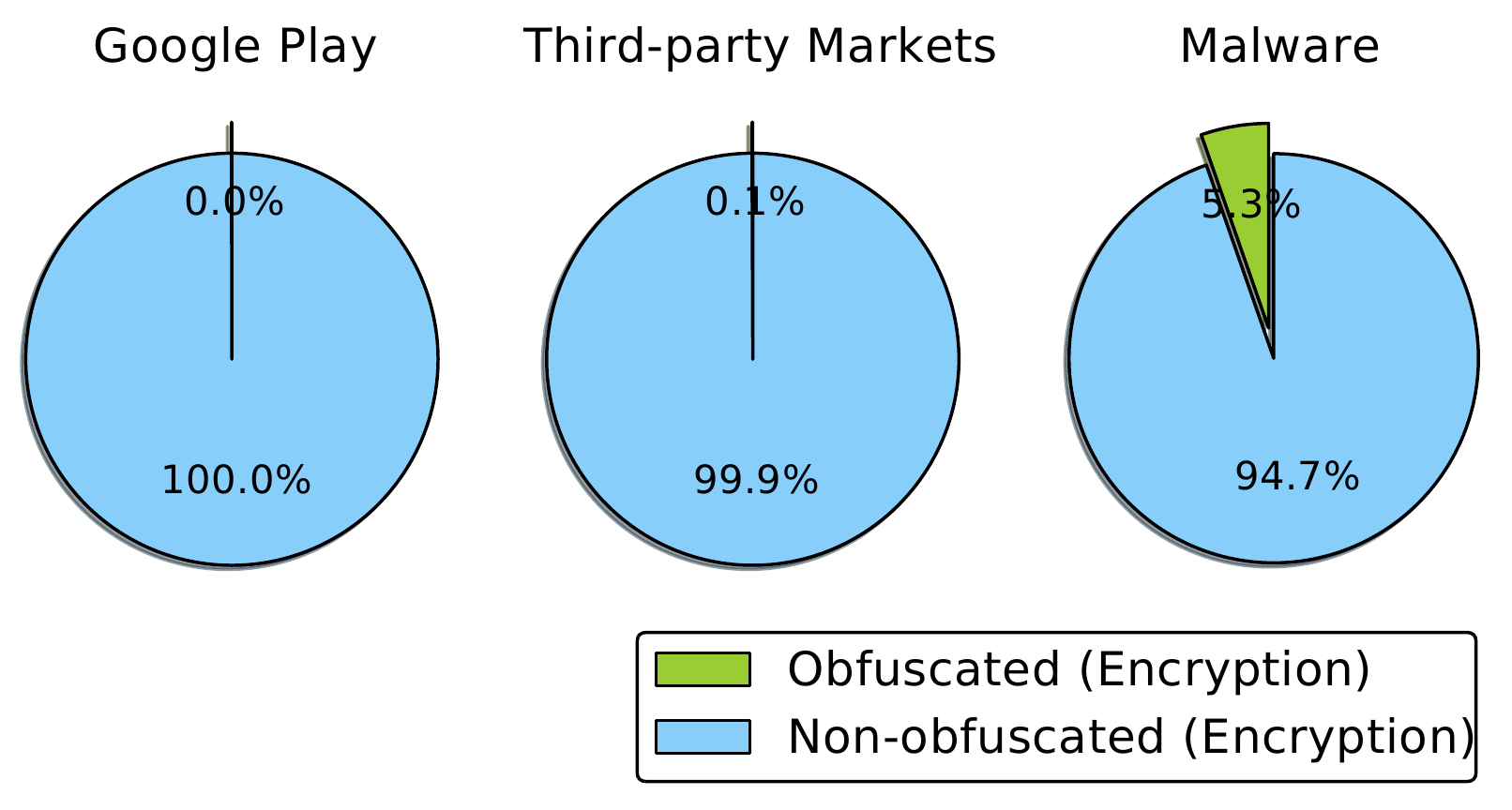}           
	\caption{Ratio of String Encryption in Three Datasets}
	\label{fig:encryption_percentage}
\end{figure} 

\begin{mybox}[boxsep=0pt,
	boxrule=1pt,
	left=4pt,
	right=4pt,
	top=4pt,
	bottom=4pt,
	]
	~ $\Rightarrow$ 1. Nearly all benign apps don't use string encryption. \\
	$\Rightarrow$ 2. String encryption is more popular in malware.
\end{mybox}

This statistical result complies with our perception, and we could understand it from two perspectives. (1) String encryption is not a common feature provided by off-the-shelf obfuscators. For example, ProGuard ~\cite{url_proguard}, as the default obfuscator integrated into Android Studio, does not provide such option. The obfuscators offering the string encryption feature are either expensive (DexGuard~\cite{url_dexguard}, DexProtector~\cite{url_dexprotector}) or difficult to configure (Allatori~\cite{url_allatori}). \ignore{Although some obfuscators provide the corresponding free trial versions (DashO~\cite{url_dasho}, Allatori~\cite{url_allatori}), users may concern about whether trial version provides complete functionalities or not.} (2) Many developers may lack the knowledge or awareness of deploying more advanced obfuscation techniques. They may believe the default identifier renaming is enough for code protection and it is not necessary to consider other techniques. (3) String encryption can help malware evade the signature scanning of some anti-virus software and hidden the intention effectively, leading to a higher rate of utilization than benign apps. 






\vspace{2pt}
In addition, we also conducted an experiment targeting at the implementations of cryptographic functions for obfuscation. In this analysis, we focused on the malware set because the other two benign datasets can not provide enough string-encrypted samples. Finally, we obtained 1,190 cryptographic functions. Base on the further reviews, we get the following findings.

\begin{mybox}[boxsep=0pt,
	boxrule=1pt,
	left=4pt,
	right=4pt,
	top=4pt,
	bottom=4pt,
	]
	~ $\Rightarrow$  The cryptographic functions usually disguise its true intention by changing to an irrelevant name.
\end{mybox}

For instance, in sample \texttt{com.solodroid.materialwallpaper\footnote{MD5: \texttt{fab2711b0b55eb980f44bfebc2c17f1f}}}, the decryption function is disguised to a common legitimate API \texttt{NavigationItem;->getDrawable()} which should be used for retrieving a drawable object. 


\begin{mybox}[boxsep=0pt,
	boxrule=1pt,
	left=4pt,
	right=4pt,
	top=4pt,
	bottom=4pt,
	]
	~ $\Rightarrow$ About 17.6\% of string-encrypted malware implement multiple cryptographic functions and take turns to use them in a single app.
\end{mybox}

In sample \texttt{com.yandex.metrica\footnote{MD5: \texttt{95f7d37a60ef6d83ae7443a3893bb246}}}, four different cryptographic functions were implemented. All of them share similar code structures -- first initializing the key, then doing the encryption/decryption. However, the key initialization procedures are quite different from each other. As a result, the workload of restoring rises significantly for analysts.


{\small
	\begin{lstlisting}[
	language={Java},
	label={lst:identifier},
	keywordstyle=\color{blue!70},
	numbers=left,                    
	numbersep=5pt, 
	xleftmargin=8pt,
	xrightmargin=5pt,
	numberstyle=\scriptsize\color{gray},
	breaklines=true,
	frame=single,
	%frame=shadowbox,
	basicstyle=\ttfamily,
	commentstyle=\color{blue} \textit,
	stringstyle=\itfamily,
	showstringspaces=false]
// In class com.yandex.metrica.impl.ad;
static final String a(String str){ 
if (c == null){
    a13840(); // key initialization function
}
    Continue ...
}
	\end{lstlisting} 
}

\begin{mybox}[boxsep=0pt,
	boxrule=1pt,
	left=4pt,
	right=4pt,
	top=4pt,
	bottom=4pt,
	]
	~ $\Rightarrow$ The secret keys used in cryptographic functions can be statically defined or dynamically generated. 
\end{mybox}

In the static case, the key is either hard-coded or directly imported as the parameter, which can be easily located and obtained. On the other hand, the dynamic key is usually generated at runtime and even could be fluctuating in different runtime context, which is nearly impossible to be handled by static analysis. The following code snippet shows an example of dynamic key generation, in which \texttt{elements[3]} is not a fixed value because of the uncertain stack trace at runtime.

{\small
	\begin{lstlisting}[
	language={Java},
	label={lst:identifier},
	keywordstyle=\color{blue!70},
	numbers=left,                    
	numbersep=5pt, 
	xleftmargin=8pt,
	xrightmargin=5pt,
	numberstyle=\scriptsize\color{gray},
	breaklines=true,
	frame=single,
	%frame=shadowbox,
	basicstyle=\ttfamily,
	commentstyle=\color{blue} \textit,
	stringstyle=\itfamily,
	showstringspaces=false]
StackTraceElement[] elements = Thread.currentThread().getStackTrace();
int hashCode = elements[3].getClassName()+elements[3].getMethodName().hashCode();
	\end{lstlisting}
}

\subsection{Reflection}

Reflection allows programs to create, modify and access an object at runtime, which brings many flexibilities. However, such dynamic feature also impedes static analysis due to those reflective invocations, especially those invoking other functions. Such uncertain behaviors could result in that the static analysis cannot capture the real intention.

In this section, we explore two questions on reflection: 

\begin{enumerate}
	\item How widespread the reflection is used in the wild? 
	\item Among these use cases, how many of them are used for the obfuscation purpose? 
\end{enumerate}

Reflection provides diverse APIs targeting at different objects like Class, Method and Field. In practice, particular APIs are often executed in sequence to achieve specific functionalities. In our study, we focus on the sequence pattern {[\texttt{Class.forName()} $\rightarrow$ \texttt{getMethod()} $\rightarrow$ \texttt{invoke()}]} which is the most frequent pattern for reflective calls mentioned by Li et al.~\cite{DBLP:conf/issta/0029BOK16}. Also, in this sequence, the execution of program is implicitly transfered to another function (the parameter of  \texttt{getMethod()}), which has an obvious influence on program status, especially the control flow. 


%


%

\vspace{2pt} \noindent
\textbf{Reflection Detection.} The first target is fast reflection detection, which could be achieved through signature searching, say the sequence pattern {[\texttt{Class.forName()} $\rightarrow$ \texttt{getMethod()} $\rightarrow$ \texttt{invoke()}]}.

Another target is to discover the invoked function in reflection, that is the input parameter of reflective calls. In theory, dynamic analysis is the best way to find the input parameter. However, its low path coverage and efficiency issues are not suitable for large-scale scanning. To balance the efficiency and coverage, we developed a light-weight tool to trace the input parameters of \texttt{Class.forName()} and \texttt{getMethod()}. The high-level idea is to find the real content of the parameters through backward slicing.

More details, first our tool scans the function body and locates two reflection calls -- {\texttt{Class.forName()}} and {\texttt{getMethod()}}. The parameter registers will be set as \emph{slicing criterion}. Then it traces back from the locations, analyzing each instruction to find the corresponding \emph{slices}. After that, this tool parses and simulates each instruction in \emph{slices}, and calculates the final value of the \emph{slicing criterion}. Note that, to reduce the maintenance complexity, we do not carry out recursive function invoking resolution.




Here, we use a real-world example (see the below code block) to illustrate such work flow. In this case, our tool will mark the positions of \emph{blue-highlighted} reflective calls and trace the data flow of \emph{red-highlighted} registers. The final output would be \{"{\texttt{android.os.SystemProperties}}", "{\texttt{get}}"\}.

			
			
				
				
				
			

	

{\small
	\begin{lstlisting}[
	language={Java},
    label={lst:xxx},
	keywordstyle=\color{blue!70},
	numbers=left,                    
	numbersep=5pt, 
	xleftmargin=8pt,
	xrightmargin=5pt,
    numberstyle=\scriptsize\color{gray},
	breaklines=true,
	frame=single,
	basicstyle=\ttfamily, 
	commentstyle=\color{blue} \textit,
	stringstyle=\ttfamily, 
	showstringspaces=false,
	mathescape=true]
$\textbf{const/4}$  $\textbf{v1}$, 0
$\color{blue!70}\textbf{const-string}$  $\color{red!70}\textbf{v0}$,'android.os.SystemProperties'
$\color{blue!70}\textbf{invoke-static}$ $\color{red!70}\textbf{v0}$,Ljava/lang/Class;->forName(Ljava/lang/String;)Ljava/lang/Class;
$\color{blue!70}\textbf{const-string}$ $\color{red!70}\textbf{v2}$, 'get'
$\dots$
$\color{blue!70}\textbf{invoke-virtual}$ $\color{red!70}\textbf{v0}$, $\color{blue!70}\color{red!70}\textbf{v2}$, $\textbf{v3}$, Ljava/lang/Class;->getMethod(Ljava/lang/String; [Ljava/lang/Class;)Ljava/lang/reflect/Method;
	\end{lstlisting} 
}

\vspace{2pt} \noindent
\textbf{Large-scale Investigation and Findings.} The implementation of our detection models (reflection usage and invoked functions in reflection) is still based on Androguard with around 1600 Python lines of code. After experiments on our APK dataset, the reflection statistics are shown in Figure~\ref{fig:reflection_percentage}. We could find:

\begin{figure}[t]
	\centering
	\includegraphics[width=0.95\columnwidth]{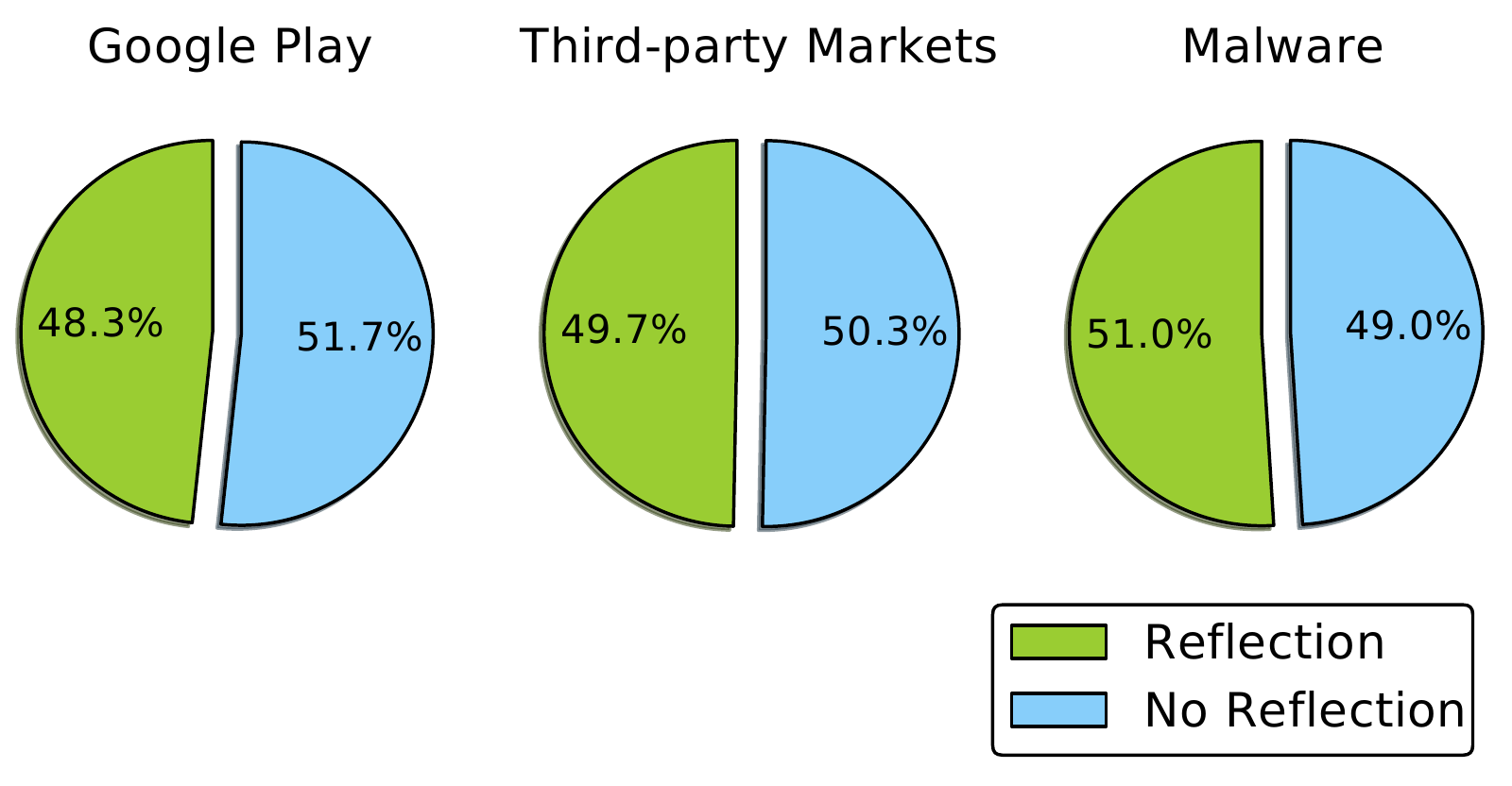}          
	\caption{Ratio of Reflection in Three Datasets}
	\label{fig:reflection_percentage}
\end{figure} 

\begin{mybox}[boxsep=0pt,
	boxrule=1pt,
	left=4pt,
	right=4pt,
	top=4pt,
	bottom=4pt,
	]
	~ $\Rightarrow$ The proportions of reflection deployment in benign apps and malware are similar.  
\end{mybox}


We are also interested in the purposes of applying reflection in apps. Since our detection model does not work at the dynamic level, part of the invoked targets cannot be precisely acquired. To some complex invoking cases, our model will try to record relevant information as much as possible. For example, if the real target is delivered as the return value of another function, our tool will record the information of this function. The percentage of recovered targets is shown in Table~\ref{tab:reflection_complete}, which indicates malware hold the least recovery rate among the three datasets. Furthermore, we checked the results of our backward slicing prototype and found that most of the strings delivered to reflection calls in malware are the return values of certain cryptographic functions, like \texttt{Ltp5x/WGt12/StringDecoder;} $\rightarrow$ \texttt{decode(Ljava/lang/String;)Ljava/lang/String;}. 

\ignore{
\begin{mybox}[boxsep=0pt,
	boxrule=1pt,
	left=4pt,
	right=4pt,
	top=4pt,
	bottom=4pt,
	]
	~ $\Rightarrow$ Compared with benign apps, malwares use more complex reflection invoking patterns.
\end{mybox}
}


 


\begin{table}[t]
	\caption{Ratio of Recovered Targets in Reflection}
	\label{tab:reflection_complete}
	\begin{tabu} to 0.488\textwidth{X[0.5,l]X[1,c]X[1,c]X[1,c]}
		\toprule
		\textbf{Dataset} & \textbf{Google Play} & \textbf{3rd-p Markets} &  \textbf{Malware} \\
		\midrule
		Recovery & 65.7\% &50.2\% &27.1\% \\
		\bottomrule
	\end{tabu}
\end{table}


\vspace{2pt}
To the successfully recovered functions, we further explore why these reflection implementations are necessary. According to different APK dataset, the most frequently invoked functions are listed in Table~\ref{tab:ref_top_goolge}, Table~\ref{tab:ref_top_3rd}, and Table~\ref{tab:ref_top_malware} respectively. These lists reflect:


\begin{table}[t]
	\caption{Functions Invoked via Reflection (Google Play)}
	\label{tab:ref_top_goolge}
	\begin{tabu} to 0.488\textwidth{X[1,l]X[4.8,l]}
		\toprule
		\textbf{Frequency}& \textbf{Recovered Function}  \\
		\midrule
		2,275 &   \texttt{android.support.v4.content.
			LocalBroadcastManager.getInstance}  \\
		\hline \rowcolor{mygray}
        1,297 &  \texttt{android.webkit.WebView.onPause} \\
		\hline
		 1,250 &  \texttt{android.os.SystemProperties.get} \\
		\hline \rowcolor{mygray}
		821 &  \texttt{org.apache.harmony.xnet.provider.jsse .NativeCrypto.RAND\_seed}  \\
		\hline
		523 &  \texttt{com.google.android.gms.common.GooglePlay-
			ServicesUtil.isGooglePlayServicesAvailable } \\
		\bottomrule
	\end{tabu}
\end{table}

\begin{table}[t]
	\caption{Functions Invoked via Reflection (3rd-p Market)}
	\label{tab:ref_top_3rd}
	\begin{tabu} to 0.488\textwidth{X[1,l]X[4.8,l]}
	\toprule
	\textbf{Frequency}& \textbf{Recovered Function}  \\
	\midrule
	 3,859&   \texttt{android.os.SystemProperties.get}  \\
	\hline \rowcolor{mygray}
	1,800&   \texttt{android.support.v4.content.
		LocalBroadcastManager.getInstance}  \\
	\hline
	1,158 &   \texttt{org.apache.harmony.xnet.provider.jsse .NativeCrypto.RAND\_seed }\\
	\hline \rowcolor{mygray}
	721 &   \texttt{android.os.ServiceManager.getService }\\
	\hline
	613&   \texttt{android.os.Build.hasSmartBar} \\
	\bottomrule
	\end{tabu}
\end{table}

\begin{table}[t]
	\caption{Functions Invoked via Reflection (Malware)}
	\label{tab:ref_top_malware}
	\begin{tabu} to 0.488\textwidth{X[1,l]X[4.8,l]}
	\toprule
	\textbf{Frequency}& \textbf{Recovered Function}  \\
	\midrule
	2,977 &   \texttt{java.lang.String.valueOf } \\
	\hline \rowcolor{mygray}
  	2,142&   \texttt{android.telephony.gsm.SmsManager.getDefault} \\
	\hline
	687 &   \texttt{android.os.SystemProperties.get}\\
	\hline \rowcolor{mygray}
 	518 &   \texttt{java.lang.String.charAt} \\
	\hline
 	352&  \texttt{java.lang.String.equals} \\
	\bottomrule
	\end{tabu}
\end{table}


\begin{mybox}[boxsep=0pt,
	boxrule=1pt,
	left=4pt,
	right=4pt,
	top=4pt,
	bottom=4pt,
	]
	~ $\Rightarrow$ Most of the reflection cases are used to invoke hidden functions or to support backward compatibility.
\end{mybox}

In Android system, the functions related to the Android framework and OS itself are usually annotated with the label "{\texttt{@hide}}", which can only be called through reflection. In above three tables, all functions starting with \texttt{android.os.*} and \texttt{android.webkit.*} are hidden-annotated.

We also manually checked the use case of \texttt{android.v4.content. LocalBroadcastManager.getInstance}. We found that the corresponding reflective calls are usually enclosed in a \emph{try-catch} block, aiming to check the existence of particular class and handle the not-found exception. Such pattern is a programming standard recommended by Android official documents~\cite{url_ref_backcompatibility}.


\vspace{2pt}
To malware samples, we find:
\begin{mybox}[boxsep=0pt,
	boxrule=1pt,
	left=4pt,
	right=4pt,
	top=4pt,
	bottom=4pt,
	]
	~ $\Rightarrow$ Compared with benign apps, malware prefers to use more complex reflection invoking patterns to hide its intentions.
\end{mybox}


As one example, the following code block is extracted from an obfuscated malware\footnote{MD5: \texttt{7ff1b8afd22c1ed77ed70bfc04635315}}. After analysis, the function invoked by reflection could be restored as: 

{\small
	\begin{lstlisting}[
	%caption={Restored Function},
	language={Java},
	label={lst:restore_func},
	keywordstyle=\color{blue!70},
	numbers=left,                    
	numbersep=5pt, 
	xleftmargin=8pt,
	xrightmargin=5pt,
	numberstyle=\scriptsize\color{gray},
	breaklines=true,
	frame=single,
	basicstyle=\ttfamily, 
	commentstyle=\color{blue} \textit,
	stringstyle=\ttfamily, 
	showstringspaces=false,
	mathescape=true]
if (!(*@\`{o}@*).trim().toLowerCase().contains((*@\^{0}@*)("G))OCH"))) {Function Body}
	\end{lstlisting} 
}

As comparison, the original code is shown below. In this case, all string operations can be written in non-reflection forms. We could find such reflection usage makes the code structure more complicated and confusing, which enhances the effect of code obfuscation. 

{\small
	\begin{lstlisting}[
	%caption={Original Function},
	language={Java},
    label={lst:orig_func},
	keywordstyle=\color{blue!70},
	numbers=left,                    
	numbersep=5pt, 
	xleftmargin=8pt,
	xrightmargin=5pt,
	numberstyle=\scriptsize\color{gray},
	breaklines=true,
	frame=single,
	basicstyle=\ttfamily, 
	commentstyle=\color{blue} \textit,
	stringstyle=\ttfamily, 
	showstringspaces=false,
	mathescape=true]
if (!((Boolean) Class.forName("java.lang.String").getMethod("contains", new Class({CharSequence.class}).invoke(Class.forName("java.lang.String").getMethod("toLowerCase", null).invoke(Class.forName("java.lang.String").getMethod("trim", null).invoke((*@\`{o}@*), null), null), new Object[]{(*@\^{0}@*)("G))OCH")})).booleanValue()) { $\textit{Function Body}$ }
	\end{lstlisting} 
}

\subsection{Packing}

Different from previous three obfuscation techniques, packing is a kind of whole-APK-reinforcing protection, which does not aim at preventing others from understanding the code, but preventing the code from being obtained. Currently, many packing services are provided as online services and free for individual users, such as Qihoo~\cite{url_360jiagu}, ijiami~\cite{url_ijiami}, and Bangcle~\cite{url_bangcle}.

\vspace{2pt} \noindent
\textbf{Packing Detection}. Our study shows the apps using packing usually have the following heuristic features:
\begin{enumerate}
	\item{\textit{Derived Application Class}}. \texttt{android.app.Application} is the base class maintaining the global app state. When launching an app, this class (or its subclass) will be instantiated first. The operation of packing apps usually needs a derived \texttt{Application} class acting as the wrapper, preparing for the subsequent APK loading.
	
	\item{\textit{Encrypted Data File}}. The real APK is usually encrypted and stored in the \texttt{lib} or \texttt{assets} folder.
	
	\item{\textit{Thin Wrapper Class}}. In general, the wrapper class only performs the bootstrap function, and the core work is performed by native functions based on Java Native Interface (JNI)~\cite{url_jni}.
\end{enumerate}

Also, the packing tools always introduce new files (such as \texttt{ijiami.data} and \texttt{baiduproduct.jar}) or code into the original APK file. These modifications usually differ from one packing service to another and can be the fingerprints of service providers. Those certain modifications could be treated as a detection feature as well. To further study, we tested six popular packing services and analyzed the corresponding packed APK files. The extracted signatures are listed in Table~\ref{tab:sig_packing}. Noted that, such signatures may be changed with the update of packing service.



\begin{table*}[t]
	\caption{Signatures of Packing Services}
	\label{tab:sig_packing}
	\begin{tabu} to 0.98\textwidth{X[1,l]X[6,l]X[4.8,l]}
			\toprule
\textbf{Packer} &  \textbf{File Signature(s)} & \textbf{Code Signature(s)} \\
\midrule
Ali & \texttt{lib/armeabi/libmobisec.so} | \texttt{aliprotect.dat} & \texttt{com.ali.fixHelper} | \texttt{com.ali.mobisecenhance.StubApplication}\\
\hline
Tencent &\texttt{lib/armeabi/libmain.so} | \texttt{lib/armeabi/libshell.so} | \texttt{lib/armeabi/mix.dex}
& \texttt{com.tencent.StubShell}\\
\hline
Qihoo & \texttt{assets/libjiagu.so} & \texttt{com.qihoo.util.StubApplication}\\
\hline
iJiami & \texttt{assets/ijiami.dat} | \texttt{*/armeabi/libexec.so} | \texttt{*/armeabi/libexecmain.so}& \texttt{com.shell.SuperApplication}\\
\hline
Bangcle & \texttt{assets/bangcle\_classes.jar} | \texttt{lib/armeabi/libsecexe.so} | \texttt{lib/armeabi/libsecmain.so}& \texttt{com.secshell.shellwrapper.SecAppWrapper}
| \texttt{com.bangcle.protect.ApplicationWrapper}\\
\hline
Baidu & \texttt{assets/baiduprotect.jar} | 
\texttt{lib/armeabi/libbaiduprotect.so} & \texttt{com.baidu.protect.StubApplication}\\
\bottomrule
	\end{tabu}
\end{table*}

\vspace{2pt} \noindent
\textbf{Large-scale Investigation and Findings}. We applied our packing detection prototype (300 Python lines of code) to the three APK datasets. The statistical results are shown in Figure~\ref{fig:packing_percentage}. The direct finding is:

\begin{figure}[t]
	\centering
	\includegraphics[width=0.95\columnwidth]{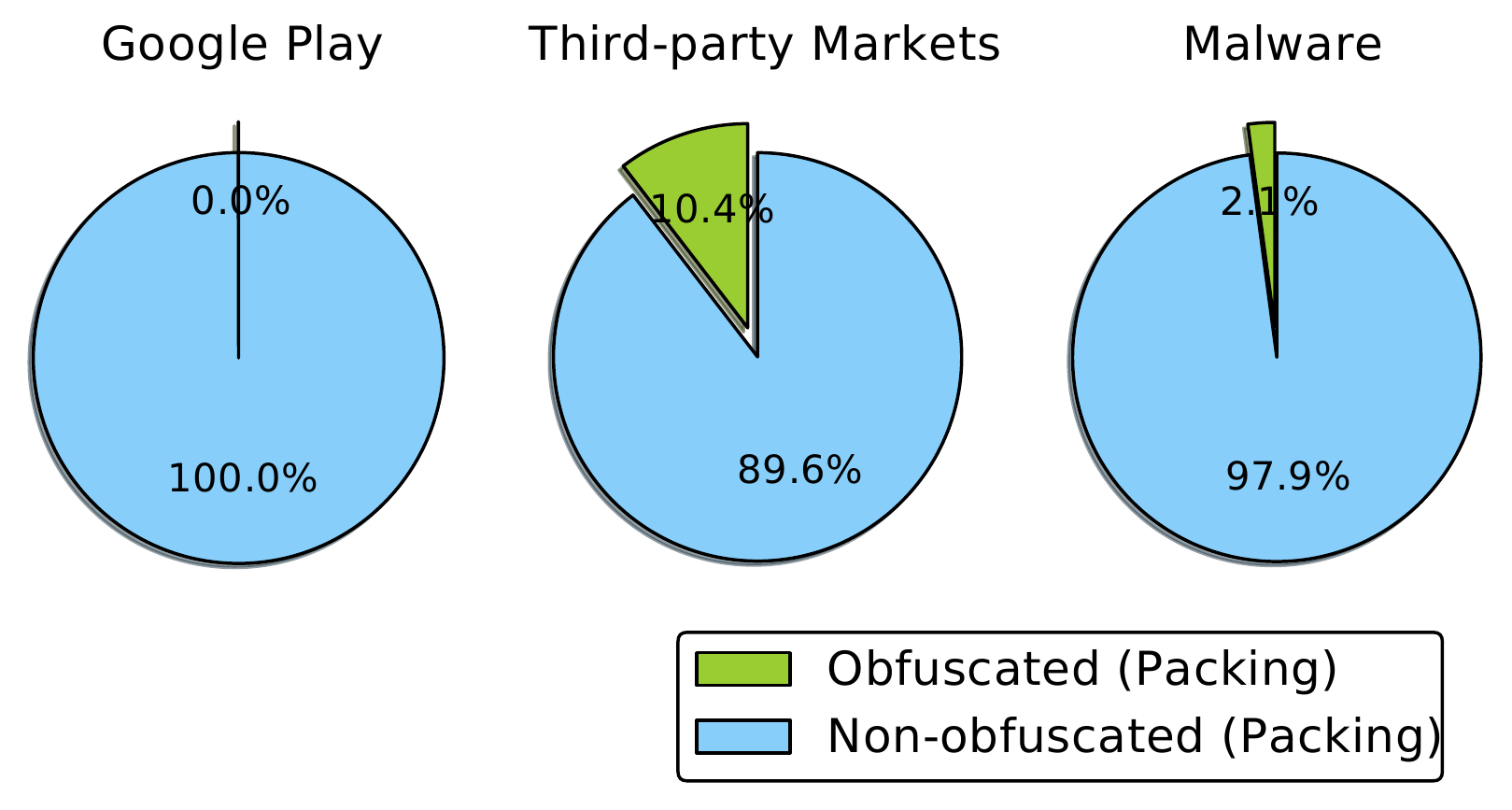}           
	\caption{Ratio of Packing in Three Datasets}
	\label{fig:packing_percentage}
\end{figure} 

\begin{mybox}[boxsep=0pt,
	boxrule=1pt,
	left=4pt,
	right=4pt,
	top=4pt,
	bottom=4pt,
	]
	~ $\Rightarrow$ Third-party apps and malware held a higher deployment rate of packing services.
\end{mybox}

As an one-stop approach of code protection, the popularity of online packing service is reasonable. Currently, the research on packing and unpacking has become a hot topic, and researchers have proposed several tools targeting at unpacking apps automatically, like Zhang et al.~\cite{DBLP:conf/esorics/ZhangLY15} and Yang et al.~\cite{DBLP:conf/raid/YangZLSLHG15}. Most of these tools rely on dumping the code from memory through customized Dalvik virtual machine (DVM) or Android Runtime (ART). As arm races, packing providers enhance their services time to time to prevent cracking.

According to our observation, packing is a practical approach to code protection for ordinary developers. Its basic functionality has been able to impede entry-level reverse-engineers from peeping into the original code. However, the protection may be not strong enough to prevent an adept analyst from obtaining the code.

\section{Discussion}
\label{sec:discussion}

In this section, we discuss some limitations of our study and then describe the future plan. Though we have conducted a large-scale investigation of mainstream obfuscation techniques used in Android apps, we should point out there are still some existing techniques not involved in our research, say control flow obfuscation and native code obfuscation. 

Through our observation, we find that control flow obfuscation is non-universal and only provided by a minority of obfuscators, like DashO and Allatori. However, based on our analysis, these tools do not provide a strong control flow obfuscation method as they claimed. For example, given an app, only very few methods' control flows are obfuscated, and the others remain unchanged. Therefore, at this stage, we cannot capture enough control-flow obfuscated samples for investigation.

Another uncovered topic is native code obfuscation which could bring more protection to an app's binary code. However, native code programming requires more advanced skills for developers, which makes it still not a mainstream technique in Android app development. Also, the implementation of native code obfuscation is quite different from other Java-level techniques, which could be treated as an independent research topic. Therefore, we leave it as our future study.

\section{Related Work}
\label{sec:related}
Obfuscation is always a hot research topic in Android ecosystem, and there are several studies performed on how to obfuscate Android apps effectively and how to measure the obfuscation effectiveness.

\subsection{Obfuscation Measurement and Assessment}
Obfuscation techniques have been widely used in the Android app development. Naturally, in academia, researchers are interested in whether these techniques do work. An early attempt is \cite{DBLP:conf/securecomm/FreilingPZ14} which empirically evaluates a set of 7 obfuscation methods on 240 APKs. Also, Park et al.~\cite{DBLP:journals/jowua/ParkKJCHP15} empirically analyzed the effects of code obfuscation on Android app similarity analysis. Recently, Faruki et al.~\cite{DBLP:journals/corr/FarukiFLCG16} conducted a survey to review the mainstream Android code obfuscation and protection techniques. However, they concentrated on the technical analysis to evaluate different techniques, not like our work based on a large-scale dataset. They show that many obfuscation methods are idempotent or monotonous. Wang et al.~\cite{DBLP:conf/icse/WangR17} defined the obfuscator identification problem for Android and proposed a solution based on machine learning techniques. The experiments indicated that their approach could achieve about 97\% accuracy to identify ProGuard, Allatori, DashO, Legu, and Bangcle. On the aspect of deobfuscation research, Bichsel et al.~\cite{DBLP:conf/ccs/BichselRTV16} proposed a structured prediction approach for performing probabilistic layout deobfuscation of Android APKs and implemented a scalable
probabilistic system called DeGuard.

Different from above research, our work is based on a large Android app datasets which cover official Google play store, third-party Android markets, and update-to-date malware families. We attempt to understand the distribution of Android obfuscation techniques and provide the up-to-date knowledge about app protection.

\subsection{Security Impact of Android Obfuscation}
As discussed earlier, the obfuscation will create barriers for Android program analysis. Works on clone / repackage detection~\cite{DBLP:conf/wisec/ZhangHZW014, DBLP:conf/issta/WangGMC15, DBLP:conf/codaspy/ZhouZJN12, DBLP:conf/essos/GuanHLZ16, DBLP:journals/tr/MingZWLZ16} find that obfuscations can impair detection results. 

Studies of malware detection also showed that obfuscation is an obstacle to malware analysis. Rastogi et al.~\cite{DBLP:conf/ccs/RastogiCJ13} evaluated several commercial mobile anti-malware products for Android and tested how resistant they are against various common obfuscation techniques. Their experiment result showed anti-malware tools make little effort to provide transformation-resilient detection (in the year 2013). After that, Maiorca et al.~\cite{DBLP:journals/compsec/MaiorcaACAG15} conducted a large-scale experiment in which the detection performance of anti-malware solutions are tested against malware samples under different obfuscation strategies. Their results showed the improvement of anti-malware engines in recent years. Recently, Hoffmann et al.~\cite{DBLP:conf/codaspy/HoffmannRMWGH16} developed a framework for automated obfuscation, which implemented fine-grained obfuscation strategies and could be used as test benches for evaluating analysis tools. Similar works are also completed by Preda et al.~\cite{DBLP:journals/virology/PredaM17}, Pomilia~\cite{pomilia2016study}, and Faruki et al.~\cite{DBLP:conf/trustcom/FarukiBLGCR14}. To handle obfuscated samples, Suarez-Tangil et al.~\cite{DBLP:conf/codaspy/Suarez-TangilDA17} propose DroidSieve, an Android malware classifier based on static analysis and deep inspection that is resilient to obfuscation.   

For malware detection, researchers mainly discussed arms race between obfuscation and malware detection. Although some malware detection tools claim to still work well in the presence of obfuscation, none could eliminate the obfuscation effects in their experimental evaluation. Our study focuses on the empirical study of security impacts of obfuscation in the wild from different views, which are complementary to existing works. That is, we statistically evaluate the distribution of obfuscation methods from views of different markets, hardening capability of obfuscations and temporal evolution, with a light-weight and scalable obfuscation detection framework. We believe some of our findings would be useful for developers and researchers to better understand the usage of obfuscation, for example, keeping pace with the development of obfuscation technique.

\section{Conclusion}
\label{sec:conclusion}

In this paper, we concentrate on exploring the current deployment status of Android code obfuscation in the wild. For this target, we developed specific detection tools for four common obfuscation techniques and performed a large-scale scanning on three representative APK datasets. The results show that, to different techniques and app categories, the status of code obfuscation differs in many aspects. For example, the basic renaming obfuscation has become widely-used among Chinese third-party market developers, while still not pervasive in Google Play market. Besides, malware authors put great efforts on more advanced code protection skills. Also, we provide the corresponding illustrations to enlighten developers to select the most suitable code protection methodologies and help researchers improve code analysis systems in the right direction.



\balance
\bibliographystyle{ACM-Reference-Format}
\bibliography{ref}

\end{document}